\documentclass[10pt,osajnl2,showpacs,letterpaper,twocolumn,floatfix]{revtex4-1}  
\usepackage{graphicx}

\begin{document}

\title{Optical vortex filtering for the detection of Electromagnetically Induced Transparency}

\author{Nathaniel B. Phillips, Gleb V. Romanov, William F. Ames, Irina Novikova}
\affiliation{Department of Physics, The College of William and Mary, Williamsburg, VA 23187, USA}

\begin{abstract}
We report the realization of an optical filter based on an optical vortex mask designed to exclusively detect a weak
coherent laser field in the presence of much stronger spatially-overlapping field.  We demonstrate the
performance of such an optical vortex filter to eliminate the strong control field and detect only a weak optical
field's transmission under the conditions of electromagnetically induced transparency. The attractive feature of
such filter is its insensitivity to optical field frequencies and polarizations, which makes it applicable for a
wide range of coherent processes.

\end{abstract}


\date{\today}

\maketitle 

\begin{figure}[h] 
\includegraphics[width=1.0\columnwidth]{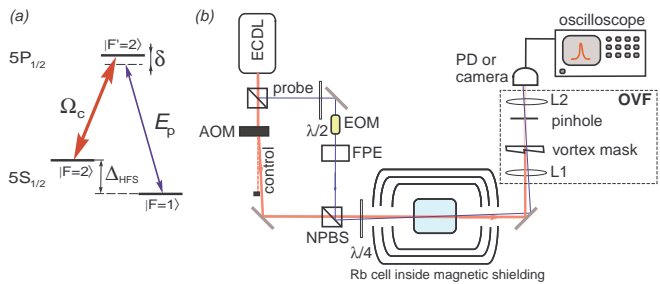}
\caption{(a) A three-level $\Lambda$-type system under conditions of EIT, where $\Omega_C$ is a strong control field,
$E_p$ is a weak probe field, $\Delta_\mathrm{HFS}$ is a hyperfine splitting between two ground state sublevels, and
$\delta$ is a two-photon Raman detuning.
(b) Schematic of the experimental setup. See text for abbreviations. \label{Fig:setup} } 
\end{figure}

The problem of detecting a weak optical signal in the presence of other strong laser fields is inherent to many
optical experiments. In particular, this problem arises in experiments involving electromagnetically induced
transparency (EIT)~\cite{Ref:EITReviewArticle,lukin03rmp}, in which optical properties of a weak probe field can be coherently controlled by means of a strong classical control field. The most common EIT configuration is a three-level $\Lambda$ system, shown in Fig.~\ref{Fig:setup}(a), in which the control and probe fields connect two long-lived hyperfine or Zeeman sublevels of the electronic ground state with a common excited level. In this case, the presence of the control field leads to a strong coupling between the probe field and a collective atomic spin excitation that results in strong suppression of resonant absorption, and opens a narrow window of transparency, $\Gamma_{\rm
EIT}=\gamma_0+|\Omega_C|^2/(\sqrt{d}\gamma)$, where $\gamma_0$ is the inverse
lifetime of the Zeeman coherence, $\Omega_C$ is the Rabi frequency associated with the control field, and $\gamma$ is the optical polarization decay rate.  Here, we define the optical depth, $2d$, such that the probe intensity without EIT is attenuated by $e^{-2d}$. The accompanying steep
normal dispersion allows for
dramatic group velocity reduction $v_g/L\propto |\Omega_C|^2/(d \gamma)\ll c$ (``slow light''), where $L$ is the length of the medium, and the reversible mapping of a probe pulse onto a long-lived
atomic coherence (``stored light'')~\cite{lukin03rmp,matskoAdvAMOP01,boydreviewJMO09}. These phenomena present promising avenues towards technologies like
miniature atomic clocks and magnetometers~\cite{vanier05apb,NISTmagnetometer}, all-optical delay lines~\cite{boydreviewJMO09},
single photon sources~\cite{kuz03etal,eisaman04,balicPRL05}, efficient quantum memories~\cite{lvovskyreviewNP09,kimbleNature08}, etc.

However, the exclusive detection of a weak probe field, especially at the few-photon level, in the presence of a strong control field becomes a challenging experimental task. Since the two optical fields have similar frequencies (differing by a few GHz for a $\Lambda$-system based of different hyperfine ground state sublevels to a sub-MHz difference for Zeeman sublevels of the same hyperfine manifold), traditional dichroic or interference filters are not effective, and thus only narrow-band transmission elements (such as high-finesse Fabry-P\'erot \'etalons) ~\cite{eisaman04,schmiedmayerPRA07,bensonRSI10} or high-quality polarizers (for orthogonal control and probe fields) must be placed before a probe detector to suppress the control field. These filtering methods are often very sensitive to frequencies and polarizations of the two optical fields, and impose restrictions on experimental arrangements.
Spatial filtering has been successfully implemented in most cold atom experiments; however,  experimental realizations of EIT in a warm atom system require  nearly collinear propagation of the signal and control fields to avoid large two-photon Doppler broadening of the EIT resonance $(\vec{k}_c-\vec{k}_p)\cdot \vec{v}_{\rm atoms}$. Thus, an appropriate spatial filter must efficiently discriminate two optical fields propagating at very small angle with respect to each other.

In this paper, we present a proof-of-principle demonstration of an optical vortex filter (OVF) that we used to detect a weak probe laser field in the presence of a strong coherent control field under conditions of
EIT.  This filtering method is largely insensitive to the polarizations and frequencies of two optical fields and thus lacks the complications associated with spectral and polarization filtering mentioned above.  It can also be used in conjunction with other filtering methods, if a single filtering method is not sufficient.

An optical vortex, also known as a phase singularity or screw dislocation~\cite{OVbook}, is a zero of optical
field intensity within an otherwise non-zero field. The cross-sectional field amplitude of a beam carrying an optical vortex can
be described, in polar coordinates $r$ and $\theta$, as:
\begin{equation} \label{OVamplitude}
E(r,\theta) \propto \left(\frac{r}{w_0}\right)^m \exp \left[-\left(\frac{r}{w_0}\right)^2\right] \exp\left[ i m
\theta\right],
\end{equation}
where $w_0$ is the beam waist, and $m$ is the topological charge of the vortex, which
characterizes the number of $2\pi$ phase variations within the vortex.

In principle, an optical vortex
beam is created by transmitting a Gaussian laser beam through a transparent phase mask having
azimuthally-varying thickness, $d=d_0-m\lambda_0\theta/(2\pi)\Delta n,$ where $d_0$ is the maximum thickness,
$\lambda_0$ is the specified wavelength, and $\Delta n$ is the refractive index difference between the mask and
its surroundings.  The beam acquires a spatial profile with an azimuthally harmonic phase $\phi=m
\theta/(2\pi)$. In the center of the mask, the phase is undefined, and the transmitted beam interferes
destructively creating a region of zero intensity.  However, due to the manufacturing difficulty of creating a smooth spiral phase of this type, a ``spiral stair-case'' mask is typically manufactured.  In our experiment, the phase plate had eight ``steps'', such that its thickness as a function of the azimuthal angle $\theta$ can be described as $d=d_0-m\lambda_0 \lfloor 8 \theta/(2 \pi)\rfloor \Delta n$, where $\lfloor \ldots \rfloor$ indicates usage of the floor function.  This phase mask produces the same destructive interference at its center---resulting in a region of zero intensity---but may produce unwanted diffraction, owing to the sharp discontinuities at the step boundaries.

The basic idea of an OVF is the following: both overlapping optical fields are focused on the vortex mask such
that the unwanted control field is aligned with its center and is converted into an optical vortex beam. The
probe field, which propagates at a small angle, does not impinge on the center of the mask, and thus is not
affected by the mask. As a result, after the vortex mask, the probe beam propagates within the dark core of the
control field beam, and the bright portion of the control field can be blocked with an appropriately-sized
aperture, leaving only the probe discernible. Similar OVFs have been effectively used to distinguish a dim
signal from a stronger background in several applications, such as the detection of extra-solar planets
\cite{GAS_OL2001,ESPtheory, ESPexperiment} and incoherent scattering \cite{palaciosPRL}.

In our experiment, we used a prototype OVF to detect the EIT resonance in a
$\Lambda$ system formed by  control and probe optical fields tuned to the
$5S_{1/2}F=2\rightarrow5P_{1/2}F'=2$ and $5S_{1/2}F=1\rightarrow5P_{1/2}F'=2$ transitions  of the D${}_{1}$ line of ${}^{87}$Rb  correspondingly. Both optical fields have the same circular polarization. The attractive feature of such polarization arrangement is that it can be accurately approximated by a simplified three-level $\Lambda$-system~\cite{phillipsPRA08}, but so far it has been limited to experiments with weak classical probe fields, since it does not allow polarization filtering.

The experimental setup for evaluation of the OVF performance is shown in Fig.~\ref{Fig:setup}(b).
Both probe and control fields were derived from the same external cavity diode laser (ECDL) to preserve their mutual phase coherence. The two beams were separated on a polarizing beam splitter. The transmitted (more powerful) beam passed through an acousto-optical modulator (AOM), and the first diffraction order with its frequency shifted by $-80$\ MHz was used as a control field in EIT experiment. The reflected beam was phase-modulated using an electro-optical modulator (EOM) at frequency $6.7547$\ GHz, which matches the hyperfine splitting of ${}^{87}$Rb ground state levels ($6.8347$\ GHz) minus $80$\ MHz. The $+1$ modulation sideband was then filtered using a temperature tuned Fabry-P\'erot \'etalon (FPE) and served as a probe field. The other EOM modulation sideband and the main carrier were suppressed by a factor of approximately $1:100$; moreover, since the modulation frequency differed from the hyperfine splitting, they do not effect probe field propagation through nonlinear mixing. The control and probe beams were then recombined in a non-polarizing beam splitter cube (NPBS) at a small angle $0.92$~mrad and
converted into parallel circular polarizations by a quarter-wave plate ($\lambda/4$) placed before the
cell. Both probe and control beams had similar divergence of approximately $0.38$~mrad, so it was virtually impossible to separate them even after long propagation. The powers of the control and probe fields before entering the Rb cell were correspondingly $1.15~$mW and
$40~\mu$W, and their respective diameters were $1.52$~mm and $0.47$~mm.
A cylindrical Pyrex cell (length $L=7.5$~cm and diameter $2.5$~cm) containing isotopically enriched ${}^{87}$Rb
and $5$~Torr of Ne buffer gas was mounted inside three-layer magnetic shielding to reduce the influence of
stray magnetic fields. The temperature of the cell was maintained at $50.0^\circ$C, which corresponded to an optical depth of $2d=49$.  Our procedure for calculating the optical depth is described in Ref.\ \cite{phillipsPRA08}.

\begin{figure}[t] 
\includegraphics[width=1.0\columnwidth]{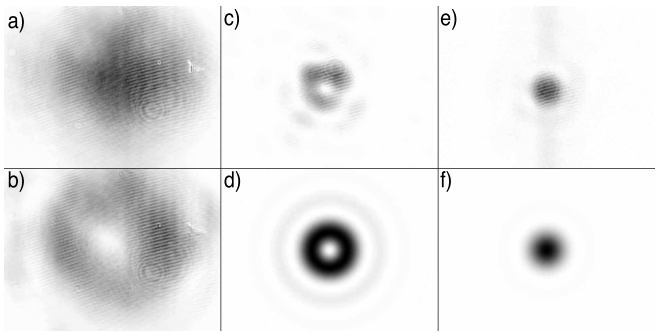}
\caption{\emph{Left column}: Measured profiles of the control field before (a) and after (b) the phase
mask. \emph{Middle column}: Recorded (c) and calculated (d) diffraction pattern of the optical vortex (the
control field after the phase mask) through the pinhole.  \emph{Right column}: Recorded (e) and calculated (f) diffraction pattern of the probe field after the phase mask through the pinhole.  All images have been inverted in order to exhibit the higher order diffraction fringes.\label{fig:beamarray}} 
\end{figure}

The OVF after the cell consisted of a bichromatic lens $\mathrm{L}1$ (focal length $f_1=35$\ mm), which focused both control and probe
beams onto a spiral mask. The vortex mask was nominally designed to produce a first-order ($m=1$) optical vortex at
$850$~nm, but it 
worked well for $795$~nm. The position of the mask was carefully adjusted such that the control field was centered at the spiral
and transformed into a vortex beam, as shown in Fig.\ \ref{fig:beamarray}(a,b). Since the probe field propagated at a
small angle, it was focused away from the vortex, and therefore its transverse intensity distribution was
practically unchanged after the mask.  A $150\ \mu$m pinhole, placed after the mask was
used to block the bright annulus of the control field while passing most of the probe field. The second lens
$\mathrm{L}2$ ($f_2=100$\ mm) collimated the beams. The transverse intensity distributions of both beams were
recorded with a commercial CCD
camera, and their total intensities were measured by a photodiode (PD).

An ideal filter will completely extinguish the control field without affecting the probe field. In reality, however, there are several factors limiting the OVF performance. For example, the vortex mask does not have a continuous thickness, but rather consists of eight steps with sharp boundaries that contribute to diffraction; also, any small manufacturing imperfections in the vortex mask center break conditions for perfect destructive interference and thus increase the amount of light in the dark part of the optical vortex beam. Also, good spatial intensity distributions for both control and probe beams is imperative, since it allows the smallest focal spot sizes at the vortex mask, and reduces diffraction losses at the pinhole.

\begin{figure}[t] 
\includegraphics[width=1.0\columnwidth]{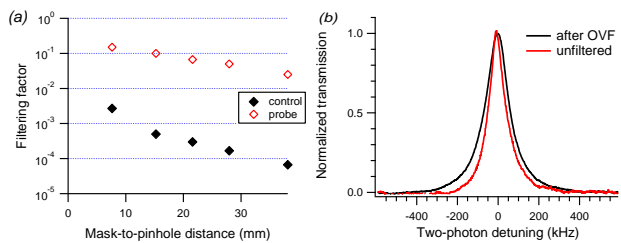}
\caption{\emph{(a)} Attenuation of the control and probe  fields after OVF. \emph{(b)} An EIT resonance in the probe field, detected after passing through the OVF (black) and without OVF (red). For easy lineshape comparison, the background is subtracted, and the resonance amplitude is normalized to one in both cases.  \label{Fig:eit} } 
\end{figure} 

To characterize OVF performance, we measured the power of the control field with and without the vortex mask and pinhole. The filtering factor, defined as the normalized power reduction of the control field, as a function of a distance between the filter mask and the pinhole, is plotted in Fig.\ \ref{Fig:eit}(a). As expected, the amount of the transmitted control light decreases when the pinhole is placed farther from the mask. Since the focus of the laser beam is at the vortex mask position, the size of the beam and the central dark spot increases, and thus less light leaks into the central area where the pinhole is placed. At the same time, even though the spatial profile of the probe field is not affected by the vortex mask, it is still attenuated by the pinhole. By performing numerical calculations of beams' propagation through OVF, we
found that the size and the position of the pinhole plays a crucial role in both control and probe
field attenuation.
While the pinhole is necessary to obstruct the bright portion of the control field, diffraction of the vortex
through the pinhole can cause the brighter outer ring to leak into the dark center and degrade the filtering
quality. Furthermore, diffraction of the signal field leads to unwanted attenuation, as energy is lost to higher
order modes and imperfect matching of the pinhole aperture size to the width of the probe beam.
%

For our experiments, the observation plane (i.e., the position of PD or camera) was approximately $z=15$\ cm from the vortex.  Thus, the diffraction criteria $a^2/(\lambda z) \approx 0.10 \ll 1$, so we are in the Fraunhofer or far-field regime.  The far-field diffraction pattern through a circular aperture of radius $a$ can be computed as follows \cite{bornandwolf, diffractionpaper}:
\begin{equation}
U(\rho,\psi)\propto\int_0^{2\pi}\int_0^a E(r,\theta) e^{\left[-i k r \rho/z \cos(\theta-\psi)\right]}r dr
d\theta,
\end{equation}
where $\rho$ and $\psi$ are the corresponding radial and azimuthal variables in the observation plane, placed at a distance $z$, and
$k=2\pi/\lambda$ is the wavenumber of light incident on the aperture.  Inserting the expression for $E(r,\theta)$
given by Eq.(\ref{OVamplitude}) with $m=1$, we find
\begin{equation}
U(\rho,\psi) \propto\int_0^{2\pi}\int_0^a \left(\frac{r}{w_0}\right) e^{-(r/w_0)^2} e^{i \theta}
e^{\left[-i k r \rho/z \cos(\theta-\psi)\right]}r dr
d\theta.
\end{equation}
The angular integral can be computed analytically \cite{gradshtein}, leaving the radial integral,
\begin{equation}\label{fraunhofer}
U(\rho,\psi)\propto\mathrm{e}^{i(\psi-\pi/2)}\int_0^a \left(\frac{r^2}{w_0}\right)\mathrm{e}^{-(r/w_0)^2}
J_1(k r \rho/z) dr,
\end{equation}
which can be evaluated via infinite series~\cite{diffractionpaper} or numerically.  Here, $J_1$ is the first order Bessel function of the first kind.  Similar methods can be used
to calculate the diffraction pattern formed by a Gaussian probe pulse through the aperture.
Figs.~\ref{fig:beamarray}(d,f) depict the results from numerically integrating the Fraunhofer diffraction
integral for both the control and probe fields. Clearly, they are in a good qualitative agreement with
experimental observations shown in Figs.~\ref{fig:beamarray}(c,e).  We show the inverted images to make the weaker diffraction fringes more discernible.  Such calculations can be used to determine the
optimal size of the pinhole that simultaneously yields maximum suppression of the control field and maximum
transmission of the probe field. Additionally, the calculations confirm that the probe inefficiency is due in large part to a non-optimal aperture size.


Finally, Fig.~\ref{Fig:eit}(b) shows the measured probe EIT resonance recorded with and without OV filtering (black and red curves, respectively). Although OV filtering does reduce the signal power due to a
non-optimal pinhole size, the method clearly preserves the overall EIT lineshape. The small discrepancy between the EIT resonance width is caused by the clipping of the wings of the intensity distribution of the probe field. In this case, we effectively detect only the part of the beam with highest and most uniform control field intensity, which results in a more symmetric power-broadened resonance. Without filtering, the regions with lower intensity (hence smaller power broadening) also contribute resulting in a overall slightly narrower resonance with sharper top~\cite{TYlineshape}. 

In conclusion, we demonstrated the possibility of using an optical vortex filter to detect a weak coherent probe field in the
presence of a much stronger control field under conditions of electromagnetically induced transparency. The best demonstrated control field
suppression factor is better than $10^{-4}$. The main limitation of the proposed method---due to the diffraction though the pinhole---can
potentially be improved by carefully selecting the pinhole size. Filtering improvement can be made by using a
phase mask with higher topological order~\cite{GAS_OL2001}. The proposed method may be beneficial for a wide
range of optical experiments due to its several advantages, such as weak sensitivity to the optical fields'
polarization and frequency, as well as only a weak distortion of the probe field wavefront.

The authors thank Kelly A. Kluttz and Carlos Lopez-Mariscal for useful discussions, and G.~A. Swartzlander for valuable advice and for loan of one of his phase masks. This research was supported by NSF grant PHY-0758010, and by the College of
William and Mary through the summer REU program.




\end{document}